\documentstyle[twoside,fleqn,espcrc2]{article}

\newcommand{\AmS}{{\protect\the\textfont2
  A\kern-.1667em\lower.5ex\hbox{M}\kern-.125emS}}

\newcommand{\be}{\begin{equation}}
\newcommand{\ee}{\end{equation}}

\title{SUMMARY TALK AT THE 3$^{rd}$ KEK TOPICAL CONFERENCE ON CP VIOLATION}

\author{R. D. Peccei
\vskip 0.1in
        Department of Physics,
        University of California, Los Angeles \\
        Los Angeles, California 90024-1547}

\begin{document}

\begin{abstract}
A summary of the contributions to this topical conference is presented.
The topics discussed ranged from detailing what we know about CP
violation, to what we
hope to learn in the future, to still unsolved mysteries in the subject.
\end{abstract}
\maketitle

\section{Introduction}
Reflecting on how to summarize the many interesting and lively contributions
made in the 3$^{rd}$ KEK Topical Conference on CP Violation, I decided to
divide my remarks into three categories:

\begin{enumerate}

\item[i)] What do we know about CP violation
\item[ii)] What are we likely to learn in the future
\item[iii)] The mysteries.
\end{enumerate}

I believe both the subject matter and the various talks we heard during this
Topical Conference nicely fit in one of these categories.
Furthermore, by proceeding in this way it allows me to follow the well
established tradition of scientific discourse, where one starts on
solid ground, proceeds to a more speculative level and ends by discussing
things in which one has no clear understanding at all!

\section{\bf What do we know about CP violation}

The study of CP violation is now 30 years old and thus
certainly is a mature subject.
Nevertheless, we still have very limited experimental information
information on this phenomena:

\begin{enumerate}
\item[i)] Our only positive evidence for CP violation is in the neutral
Kaon system, where we have measurements of two complex ratios
$\eta_{+-}$ and $\eta_{oo}$ and of the semileptonic $K_L$ symmetry,
$A_{K_{L}}$.

\item[ii)]  We have, however, a variety of bounds on possible other CP
violating phenomena, most notably very stringent  bounds on the electric
dipole moment of the neutron and of the electron \cite{PDG}.

\item[iii)]  Furthermore, one can  adduce evidence from astrophysics
and cosmology that CP violating phenomena played an important
role in the evolution of our universe.
In particular, the ratio of baryon to photons in the universe now
$n_B/n\gamma \sim 4 \times 10^{-10}$ \cite{asymmetry} is a measure of CP
violation, if the baryon asymmetry in the universe is generated
dynamically \cite{Sakharov}.
\end{enumerate}

Both the evidence for CP violation in the neutral Kaon system, as well
as the bounds on the neutron and electron dipole moments,
are consistent with
the Cabibbo Kobayashi Maskawa (CKM) paradigm \cite{CKM},
where CP violation
arises as a result of having a complex quark mixing matrix.
It is likely, however, that the baryon to photon ratio is connected with
a different source of CP violation, lying beyond the standard model.
I would like to expand on the first of these statements here and I will
return to the second point towards the end of this talk.

Qualitatively the measured CP violating parameters in the neutral
K-system can be summarized by the following observations:

\begin{enumerate}
\item[i)]  The dominant source of CP violation comes as a result of
$K-\bar K$ mixing $(\Delta S = 2$ CP violation).  Thus
\[
\eta_{+-} = \epsilon + \epsilon' \simeq \eta_{oo} = \epsilon - 2
\epsilon'~~,
\]
since the $\Delta S = 1$ CP-violating parameter $\epsilon '$ is
much smaller than the $\Delta S = 2$ CP violating parameter
$\epsilon$.

\item[ii)]  The width matrix in the $K-\bar K$ system is dominated by
the $2\pi$ intermediate state and CPT is conserved.
This circumstance leads to the following two relations \cite{Cronin},
which are well satisfied experimentally
\[
A_{K_{L}} \simeq 2 Re \eta_{+-}~,
\]
\[
 \phi_{+-} \simeq \phi_{SW} =
\tan^{-1} \frac{2\Delta m}{\Gamma_{S}-\Gamma_{L} }
\simeq 45^o~.
\]
\end{enumerate}

The CKM paradigm provides a semiquantitative explanation of the first of
these observations, while (assuming CPT conservation) the second set of
results follows essentially kinematically, given the strong $\Delta I=
1/2$ enhancement of the $2\pi$ channel.
Indeed, this enhancement is also partly responsible for the reason why CP
violation due to mixing is much greater than that due to
$\Delta S =1$ processes (direct CP-violation).
However $\epsilon ' \ll \epsilon$ follows also because the $\Delta S
= 1$ processes are Zweig-rule violating processes, involving
electromagnetic or  gluonic Penguin contributions \cite {GP}.
Schematically, one finds in the CKM model that \cite{Puri}
\[
\frac{\epsilon '}{\epsilon} \sim \frac{ReA_{2}} {ReA_{0}} \left
[\frac{\alpha_s}{12\pi} \ell n~ m_{t}^2/m_{c}^2 \right ] \sim
10^{-3}\quad ,
\]
where the first factor above contributes the $\Delta I = 3/2$ to
$\Delta I = 1/2$ suppression of about 1/20.
Furthermore, the CKM model also provides a qualitative
understanding of why
$\epsilon$ is of $O(10^{-3})$ without having to appeal to the presence
of a small CP violating phase.
Again, qualitatively, one has in the case of 3 generations \cite{Puri}

\[
\epsilon \sim \frac{Im M_{12} }{Re M_{12}} \sim
\frac{s_{12} s_{23} s_{13} \sin\delta}{s_{12}^2} \sim
10^{-3} \sin \delta~.
\]
That is, $\epsilon$ is small because the product of mixing angles
entering in the box graph contributing to this parameter is already
experimentally of $O(10^{-3})$, independent of what the phase $\delta$
is.

It turns out that the CKM model also provides a qualitative
explanation for
why no electric dipole moment for the neutron has been observed.
It is easy to convince oneself that there is no one-loop contribution to
the edm, since all the phase information cancels.
Remarkably, \cite{Shabalin} in the CKM model also the sum of all
two-loop graphs cancels, so that the first non-trivial contribution
arises at the three-loop level.
This contribution can be estimated to be of order \cite{Puri}

\[
d_n \sim e~m_{d} \frac{\alpha^2\alpha_{s} }{\pi^3}
\frac{m_{t}^2 m_{b}^2}{M_{W}^6}
s_{12}s_{23}s_{13}\sin\delta \sim 10^{-32} ecm
\]
which is many orders of magnitude below the present bound \cite
{PDG}.

Since the experimental value of $\epsilon$ (or
$\eta_{+-})$ just fixes the value of the CP phase $\delta$-assuming the
mixing angles are known precisely (see below for a more detailed
discussion) --
the value of $\epsilon '$ is the only quantitative test of the CKM model
available at present.
However, as we learned from the talk of Tschirhart, \cite {Tschirhart}
the experimental situation remains inconclusive, because the CERN and
FNAL experiments do not quite agree on what $\epsilon '$ is.
One has
\[
Re \frac{\epsilon'}{\epsilon} =
\left \{\begin{array} {ll}
(23\pm 7)\times 10^{-4} & [{\rm NA}31] \\
(7.4 \pm 5.9) \times 10^{-4} & [{\rm E}731]
\end{array}
\right.
\]
Futhermore, as Reina \cite{Reina} showed, the theoretical calculation of
this ratio still has rather large hadronic matrix element uncertainties,
and also suffers from a lack of accuracy in the knowledge of the
values of the relevant CKM matrix elements.
Although the best theoretical analysis of the problem seems to favor
the lower Fermilab value (see below), it is really too early to make
a definite pronouncement on this score.

Irrespective of theoretical prejudices, it is important to resolve in
the Kaon system the experimental controversy regarding
$\epsilon '/\epsilon$.
A clear measurement of a non zero value for $\epsilon '$ would provide
the first proof of the existence of direct CP violation -- something
that if one believes in the CKM paradigm one knows  must exist!
A measurement of $\epsilon '/\epsilon$ to the level of $10^{-4}$ should
emerge in the next round of CP violation experiments at CERN and
Fermilab \cite{Tschirhart} and possibly from the Frascati $\Phi$
Factory \cite{Bertolucci}.
Other precision experiments using both  charged and neutral Kaons,
which are now

\begin{table*}[hbt]
\newlength{\digitwidth}\settowidth{\digitwidth}{\rm 0}
\caption{Prospects and Expectations for CP Violation Tests in
Rare Decays \protect\cite{WW}}.
\label{tab:ground}
\begin{tabular}{c c c c c}
\hline
Process & \mbox{ } &  CKM Expectations  & \mbox{ } &
Experimental Prospects  \\ [1mm]
\hline
$K_S \rightarrow 3\pi^0$ & & $\epsilon^\prime_{000}/\epsilon \sim 10^{-2}$&
& $\delta \eta_{000} \sim 4 \times 10^{-3}$ \\ [1mm]
$\eta_{000} = \epsilon + \epsilon^\prime_{000}$ & & & &
[$\Phi$ Factory] \\ [1mm]
\hline
$K^\pm\rightarrow \pi^\pm \pi^\mp \pi^\pm$ & & $\Delta\Gamma \leq 10^{-6}$
& & $\delta(\Delta\Gamma)\sim 5\times 10^{-5}$ \\ [1mm]
$\Delta\Gamma=\frac{\Gamma_+-\Gamma_-}{\Gamma_++\Gamma_-}$ &  &
$\Delta g \leq 10^{-4}$ & & $\delta(\Delta g)\sim 5\times 10^{-4}$ \\  [1mm]
$\Delta g=\frac{g_+-g_-}{g_++g_-}$ & &  & & [$\Phi$ Factory] \\  [1mm]
\hline
$K^\pm\rightarrow \pi^\pm \pi^0\gamma$ & & $\Delta\Gamma \sim 10^{-3}-10^{-5}$
& & $\delta(\Delta\Gamma)\sim 2\times 10^{-3}$  \\ [1mm]
$\Delta\Gamma=\frac{\Gamma_+-\Gamma_-}{\Gamma_++\Gamma_-}$ & &    & &
[$\Phi$ Factory] \\ [1mm]
\hline
$K_L\rightarrow \pi^0 \ell^+\ell^-$ & & $B_{\rm direct} \sim 10^{-11}-10^{-13}$
& & $B\leq 7\times 10^{-11}$ \\  [1mm]
$B(K_L\rightarrow \pi^0 \ell^+\ell^-)$ & &   & & [KTeV; NA48] \\  [1mm]
\hline
$K_L\rightarrow \pi^0 \nu \bar\nu$ & & $B\sim 10^{-12}$ &  &
$B\sim 10^{-8}-10^{-9}$ \\ [1mm]
$B(K_L\rightarrow \pi^0\nu \bar\nu)$ & &  &  & [KTeV] \\  [1mm]
\hline
\end{tabular}
\end{table*}
\bigskip
\noindent
in the planning stage, potentially have bearing on
whether direct CP violation exists.
However, as Table 1 shows, these experiments are
unlikely to reach the required level
of sensitivity \cite {Tschirhart}\cite{Bertolucci}

I should comment that it is quite important also to try to pursue
experiments which are particularly sensitive to non-CKM sources of CP
violation.
A well-known example is offered by attempts to measure an electric
dipole moment for the neutron.
Two other nice examples were discussed in this Conference.
Shimizu\cite{Shimizu} reported on an ongoing
experiment at KEK aimed at looking
for T-violation in slow neutron capture in a polarized $^{134}La$
target.
Values of the triple correlation parameter
\[
\lambda \sim \langle \vec\sigma_n \cdot (\vec k \times \vec I_{La})\rangle
\]
down to $\lambda \sim 10^{-2}$ should be accessible
in this experiment, which in
sensitivity compares to a measurement of the neutron edm to $d_n \sim 5
\times 10^{-24}$ ecm.
A second experiment discussed here by Y. Kuno \cite{Kuno} --
and whose theoretical implications were illustrated by C. Geng
\cite{Geng} -- concerns measuring the transverse muon polarization in
charged $K$ decay $(K^+ \rightarrow \pi^o \mu^+ \nu_\mu)$.
This polarization again is proportional to a triple correlation
\[
\langle p_T\rangle \sim \vec\sigma_{\mu} \cdot (\vec p_{\pi^o}
\times \vec p_{\mu})
\]
and is sensitive to a possible effective scalar weak interactions,
such as those induced by an extended Higgs sector.
Writing this effective scalar interaction as
\[
M_{\rm eff} \simeq G_F \sin \theta_c m_K \xi \bar{\mu} (1-\gamma_5)
\nu_\mu ~,
\]
through a measurement of $\langle P_T\rangle$ the E246 experiment at
KEK should be sensitive to values of $Im\xi$ down to
$\delta Im\xi = 2 \times 10^{-3}$.
Such a measurement, particularly if there is a
light charged Higgs, is much more
sensitive to possible CP violating phases in the Higgs sector than a
direct measurement of $d_n$ \cite {Geng}.

Even though it is important to test for non CKM CP violating phenomena,
clearly a dominant theme in the study of CP violation in the future will
remain trying to ascertain how well the CKM paradigm actually works.
For these purposes, in my view, the more relevant tests will occur not
in the Kaon system but in the decay of neutral $B$'s to CP-self
conjugate final states.
To understand the expectations of the CKM model for these decays, it
will be useful for me
to enter into a bit of detail on  what is known about the CKM
matrix itself.

It is convenient to write the CKM matrix in
the Wolfenstein form \cite{LW},
where the three mixing angles are expanded in terms of powers of the
Cabibbo angle,
$\sin\theta_c \simeq \lambda \simeq 0.22$. One has \cite{LW}
\[
\sin\theta_{12} = \lambda ~~ \sin\theta_{23} = A \lambda^2 ~~
\sin\theta_{13} = A\sigma\lambda^3  \, .
\]
\noindent
As we shall see, experimentally both A and $\sigma$ are of $O(1)$.
In terms of the above, to $O(\lambda^3)$ the CKM matrix can be written
as
\[
V_{\rm CKM} =
\left|
\begin{array}{clcr}1 - \frac{\lambda^2}{2}
& \lambda & A\lambda^3 \sigma e^{-i\delta} \\
- \lambda & 1 - \frac{\lambda^2}{2} & A \lambda^2 \\
A\lambda^3 (1-\sigma e^{i\delta}) & -A\lambda^2 & 1
\end{array} \right|.
\]
Often, instead of $\sigma$ and the CP violating phase $\delta$, one uses
instead the parameters $\rho$ and $\eta$ with
\[
\sigma e^{i\delta} = \rho - i\eta \, .
\]

Experimentally $A$, which is related to $V_{cb}$, is knows to about 10\%,
while both $\sigma$ and $\delta$, or $\rho$ and $\eta$, are known to
only about 30\%.
These errors, however, are $\bf{not}$ measurement errors, but arise from
trying to extract theoretically these parameters from experiment.
These uncertainties were discussed in considerable detail in this
Conference.
The parameter $A$ is extracted from semileptonic B decays, either through
an inclusive analysis or by focusing on some particular exclusive mode.
In the former case, one needs to remove the sensitivity of the rate on
the uncertain $m_b$ mass $(\Gamma \sim m_b^5!)$.
For the exclusive case, heavy quark effective theory (HQET) determines
the form factors at zero recoil for the process $B \rightarrow D^* \ell
\nu$ \cite{Neubert}. However one needs to extrapolate the data to this
point,
which induces in general uncertainties. (For a contrary opinion,
see M. Tanaka \cite{Tanaka} in these proceedings).
Using a new average value for the B lifetime, mostly determined by LEP
data and by CDF data,

\[
\langle \tau_B\rangle = 1.535 \pm 0.025~~ ps~,
\]
M. Witherell \cite{Witherell} in this Conference concluded from a
combined fit of the Cleo semileptonic data that
\[
|V_{cb} | = 0.042 \pm 0.001 \pm 0.004 ~.
\]
Here the last error is an estimate of the model dependence of the
results.
The equivalent value obtained by extrapolating to the zero recoil point
the data for the exclusive $\bar B^o \rightarrow D^{*+} \ell^-\nu$
decay
measured by Argus and Cleo gives \cite{Witherell}
\[
|V_{cb}| = 0.040 \pm 0.006
\]
The average  of these two determinations \cite{Witherell} yields
\[
|V_{cb}| = 0.041 \pm 0.005 ~~ {\rm or} ~~
A = 0.85 \pm 0.10
\]

One can perhaps imagine reducing the above error to the 5\% level in the
future.
For the inclusive analysis, the model dependence of the results can be
alleviated since corrections to the parton model results are under
control, via a combination of a QCD operator product expansion and a
$1/m_{b}$ expansion \cite {BSUV}.
Furthermore, more data for the process $B \rightarrow D^* \ell \nu$
should allows a better extrapolation to be done in the exclusive
analysis.

This optimism, however, is not quite warranted for the case of $\sigma =
\sqrt{\rho^2+\eta^2}$.
This parameter is related to $|V_{ub}|$. First of all,
at the moment, we still do not
have any uncontroversial indications for an exclusive signal, like $B
\rightarrow \rho \ell \nu$. Furthermore,
for the inclusive case, to see evidence for $V_{ub}$ one must look very
near the kinematic end point
for the electron spectrum in
semileptonic B decays,
since beyond $P_\ell = 2.3~~ GeV$ the decay $B\rightarrow X \ell \nu$,
with X containing a charmed state, is kinematically forbidden.
Using the recent data from Cleo for leptons beyond the charm end
point, Witherell\cite{Witherell} arrives at the following two values for
$|V_{ub}|/|V_{cb}|$, depending on whether he
extracts this value from the data by using
a partonic approach
[ACM model \cite{ACM}] or whether he sums over exclusive final states
[IGSW model \cite{IGSW}]:
\[
\frac{|V_{ub}|}{|V_{cb}|} =
\lambda\sigma
= \left\{ \begin{array}{ll}
        0.076 \pm 0.008~ [ACM]\\
        0.101 \pm 0.010~ [IGSW]
\end{array}
\right.
\]
\noindent
Expanding the theoretical uncertainty somewhat, this analysis leads to
a value for the CKM matrix element ratio
\[
|V_{ub}|/|V_{cb}| = 0.085 \pm 0.025 ~~{\rm or}~~ \sigma = 0.38 \pm 0.11
\]

In this Conference, C. S. Kim \cite{Kim} presented a thorough analysis of
the model dependence of the results for $|V_{ub}|/|V_{cb}|$.
For this ratio one is really much more dependent
on models, since it is difficult
to apply the QCD operator expansion technique as one has no longer a
controlled $1/m_{b}$ expansion.
Rather, the expansion
\vbox{
\vspace{6cm}
\noindent
Fig. 1.  Allowed region in the $\rho-\eta$ plane.  Also shown in the
figure are what parameters determine the boundary of this
region.
}
\vspace{.5cm}

\noindent
parameter because of soft gluon emission, here is more
like $1/m_{b}(1-E_{e}/E_{\rm max})$ which fails
near the
end point\cite{MW}.  It is a very crucial question--and one of
much
current interest -- whether
one can avoid these difficulties somehow and obtain a more reliable
estimate for $\sigma$.

Information on the CP violating phase $\delta$ comes, of course, from
the measured value for $\epsilon$.
However, since $\delta$ also enter in $V_{td}$, one can obtain some
restriction on this parameter from the observed $B_d-\overline{B}_d$ mixing
parameter $x_d$, which is proportional to $|V_{td}|^2$.
Again, there are theoretical uncertainties here related to the poor
knowledge of the hadronic matrix element connected with $|\epsilon|$
and $|x_d|$.
The former is characterized by the, so-called, $B_K$ parameter, which
lattice computations give to a 20\% accuracy [$B_K = 0.8 \pm 0.2$
\cite{Martinelli}].
For $x_d$ the relevant parameter \cite{Puri} is an effective B-decay
coupling constant, which is again best determined through lattice
computations \cite{Martinelli}
\[
f_B^{eff} = \sqrt{B_B\eta_B} f_B = (200 \pm 35) MeV
\]

Because of these matrix element uncertainties, the rather
precisely measured values for
$|\epsilon|$ and $x_d$ give allowed bands in the $\sigma-\delta$ or
$\rho-\eta$ plane.
As an illustration, the overlap of these bands with the region allowed
by our present (imperfect) knowledge of $\sigma = \sqrt{\rho^2 + \eta^2}$
is shown in Fig. 1, assuming $m_t = 140$ GeV.
One sees from this figure that the physically allowed values for $\rho$
and $\eta$ in the CKM model are now constrained to a rather small region
in the $\rho-\eta$ plane.

I will return below to discuss the implications of Fig. 1
for CP violation in B decays.  I note here only that, for $m_t$=140 GeV,
using my guesstimate of the calculations presented by Reina\cite{Reina}
for the matrix element involved in $\epsilon^\prime/\epsilon$ one has
\[
\epsilon^\prime/\epsilon=(11\pm 4)\times 10^{-4} A^2\eta~.
\]
\noindent
Fig. 1 makes apparent that the expected value for $\epsilon'/\epsilon$
in the CKM model favors the result obtained by the Fermilab E731
experiment. \cite{Tschirhart}

\section{What are we likely to learn in the future?}

One should see considerable clarification of the nature of CP violation in
the coming years.  In the Kaon sector, as Tschirhart\cite{Tschirhart} and
Bertolucci\cite{Bertolucci} discussed, new experiments at FNAL [KTeV] and
at CERN [NA48], as well as at the Frascati $\Phi$ Factory [KLOE], should
push the error on $\epsilon^\prime/\epsilon$ to $O(10^{-4})$.
However, the next round of rare K decays which are
sensitive to the phase $\delta$ of the CKM matrix [cf. Table 1] will
only set limits rather than provide a new measurement for this phase.

Even though the efforts to uncover CP violating phenomena in the
Kaon sector are impressive, it is clear that much of the future
experimental activity will be focussed in the B sector.  After a long
struggle, not one but two asymmetric B-factories will begin construction
in 1994 and are aiming to take first data in 1998.  Although nobody at
this Conference discussed in detail the SLAC B factory, the talk by
Abe\cite{Abe} on the KEK project served to give an indication of the
general characteristics of these facilities.  The KEK B-factory will have
two new rings in the existing Tristan tunnel, with 3.5 GeV positrons and 8 GeV
electrons directly injected into these storage rings.
The beams will cross at an angle of 2.8 mrad and the machine parameters
are designed so as to achieve an initial luminosity of ${\cal{L}}=2\times
10^{33}~cm^{-2}~sec^{-1}$.  Eventually, by increasing the beam current,
at KEK one hopes to achieve ${\cal{L}}=10^{34}~cm^{-2}~sec^{-1}$.
As we shall see,
this high luminosity is necessary if one wants to insure the full reach for
measuring the CP violation asymmetries expected in the B sector in the
CKM model.

It was clear at the Conference, however, that
considerable other activity is also
going on in the world in B-physics.  Indeed, through the 1990's much will
be learned about the B system from experiments at LEP and at the Fermilab
Collider and, particularly, at the CESR ring in Cornell.  This was nicely
illustrated in the talks of Witherell\cite{Witherell} and Lockeyer
\cite{Lockeyer}.
As an example of the nice results obtained,
Witherell
showed evidence at the $10^{-5}$ BR level
for the important
2-body decay mode $B\rightarrow \pi\pi/K\pi$
measured at CESR
(the CLEO II data cannot
yet distinguish among these two possibilities, for lack of statistics).
Lockeyer showed
a clear discovery signal for the $B_s$ meson,
through the decay $B_s\rightarrow \psi\phi$, seen by CDF.  These results,
besides their intrinsic value, are also important ``engineering" information
for efforts to study CP violation and $B_s-\bar{B}_s$ mixing at both
electron and hadronic colliders.

Lockeyer's talk at the Conference
\cite{Lockeyer} made it clear that hadronic colliders can be very
competitive and complementary to the $e^+e^-$ B-factories in this
respect, because the more difficult experimental environment of hadron
machines can be compensated by the enormous rate for B production.  For
instance, with the main injector luminosity, there will be about $10^{11}$
B's produced per year at the Fermilab Collider, while at the LHC one will
have over $10^{12}$ B's/year.
Because of these high rates, a
very active program of experiments for
the year 2000 and beyond are now being considered.  Fermilab has
called for new collider expressions of interests by May 1994 and the LHC
experiments committee will be
examining three LOI for dedicated B detectors in
the same time frame.  How competitive all of these efforts will be to the
B factories will depend (besides on schedules!) on the ability to make
progress on triggering on certain modes (e.g. $B_d\rightarrow \pi^+\pi^-$)
and on B-tagging in the harsh hadronic environment.\cite{Lockeyer}

The detailed physics which will be pursued at the B factories
and in the hadronic colliders was discussed at this
Conference by H. Quinn\cite{Quinn}.  Fundamentally what
one wants to check first is whether the CKM paradigm is
correct.  This can be done best in the neutral B system
by testing the, so called, unitarity triangle\cite{triangle}.
Of course, this is not the only place where one can look for
CP violating effects in the B system.  However, both in
charged B decays--discussed in this Conference by Hou\cite{Hou}
--or in radiative B decays, like $B\rightarrow K\pi\gamma$--discussed
here by Soni\cite{Soni}--the actual expectations for CP violating
phenomena are much more dependent on hadronic dynamics and even
the cleanest cases are often rather rate limited.

The unitarity of the 3 generation CKM matrix implies for the $d-b$
piece that
\[
V_{ud}V^*_{ub}+V_{cd}V^*_{cb}+V_{td}V^*_{tb}=0
\]
or, approximately, to $O(\lambda^3)$
\[
V_{ub}^*+V_{td} \simeq \lambda V^*_{cb}=A\lambda^3~.
\]
Since $V^*_{ub}\simeq A\lambda^3~(\rho+i\eta)$ and
$V_{td}=A\lambda^3~(1-\rho-i\eta)$, one sees that the
unitarity of the CKM matrix gives one a triangle in the
$\rho-\eta$ plane.  This triangle has a base going from
$\rho=0$ to $\rho=1$ and its apex is any of the points
($\rho,\eta$) which are allowed in Fig. 1.  Remarkably,
the three angles in this ``unitary triangle"\cite{triangle}
are in principle measurable in B decays to CP self-conjugate
states\cite{selfconjugate}.  Hence these decays should
provide a direct test of the CKM paradigm.  As shown in Fig. 2
the angle $\beta$ spans a much narrower range than the
angles $\alpha$ and $\gamma$.  For example, for
$m_t$ = 140 GeV one has $16^\circ\leq \beta \leq 27^\circ$.

\vbox{\vspace{6 cm}
\noindent
Fig. 2.  Two examples of unitarity triangles allowed by the
CKM paradigm for $m_t$ = 140 GeV.
}
\vspace{.5cm}
The angles $\alpha,\beta$ and $\gamma$ in the unitarity triangle
can be extracted by measuring asymmetries in the decay of neutral B
mesons to self-conjugate final states $f$, where $\bar{f}=
\eta_ff$ with $\eta_f=\pm 1$.  As Quinn\cite{Quinn} showed, a
state born at $t=0$ as a B meson which decays into a final
state $f$, has a different time evolution than a state which at
$t=0$ was a $\bar{B}$.  One finds, under the assumption that
only one weak amplitude dominates\cite{selfconjugate}:
\begin{eqnarray*}
\Gamma~(B_{\rm phys}(t)\rightarrow f) & = & \Gamma(B\rightarrow f) \\
        & \times & e^{-\Gamma t} \{1-\eta_f \lambda_f \sin\Delta mt\}
\end{eqnarray*}
\begin{eqnarray*}
\Gamma(\bar{B}_{\rm phys}(t)\rightarrow f) & = & \Gamma(B\rightarrow f) \\
       & \times  & e^{-\Gamma t} \{1+\eta_f\lambda_f \sin\Delta mt\}
\end{eqnarray*}
Here $\lambda_f$ contains information about CP violation.  In the
CKM model $\lambda_f$ depends directly on the angles of the
unitarity triangle and one finds

\[
\lambda_f=\left\{\begin{array}{ll}
\sin~2\alpha & {\rm for}~B_d~{\rm decays~ involving~a} \\
 &~b\rightarrow u~
                                   {\rm transition} \\
& \\
\sin~2\beta & {\rm for}~B_d~{\rm decays~ involving~a}  \\
& ~b\rightarrow c~
  {\rm transition} \\
& \\
\sin~2\gamma & {\rm for}~B_s~{\rm decays~ involving~a}  \\
& ~b\rightarrow u~
  {\rm transition} \\
& \\
0 & {\rm for}~B_s~{\rm decays~ involving~a} \\
& ~b\rightarrow c~
  {\rm transition}
\end{array}
   \right.
\]

The prototype decay for measuring the angle $\beta$ is the
decay $B_d\rightarrow \psi K_S$.  Given the allowed region in
the $\rho-\eta$ plane of Fig. 1, the relevant asymmetry between
the rates of $(B_d)_{\rm phys}$ and $(\bar{B}_d)_{\rm phys}$ into this
mode is very large [e.g. for $m_t$ = 140 GeV, one has
$0.53\leq \sin~2\beta\leq 0.81$].  Furthermore, for the
decay $B_d\rightarrow \psi K_S$ the assumption of having
only one (effective) weak amplitude is well justified
\cite{LP}, since in this case both the weak transition
$b\rightarrow c\bar{c}s$ and the associated Penguin amplitude
have the same weak phase.  Thus this decay mode has excellent
theoretical prospects.  As Abe\cite{Abe} discussed in his talk,
the process $B_d\rightarrow \psi K_S$ is also fine from an
experimental point of view.  Using the leptonic decays of the
$\psi$ and the charged pion decays of the $K_S$, Abe
estimates that with an integrated luminosity of $100~fb^{-1}$ the
error in $\sin~2\beta$ at the KEK B factory should be
$\delta\sin~2\beta$
= 0.081.  Obviously, this measurement would
suffice to establish that $\sin~2\beta$ is non-vanishing, if it is
in the range expected in the CKM model.  Furthermore, this is not
the only decay of $B_d$ states
in which $sin~2\beta$ is accessible\cite{Quinn}.
Indeed, as Hall\cite{Hall} commented, it is likely that besides the Cabibbo
angle and $V_{cb}$, $\sin~2\beta$ will
eventually be the next best known weak
parameter connected to the CKM matrix.

The angle $\alpha$, whose prototype decay is $B_d\rightarrow \pi^+\pi^-$,
appears to be much harder to pin down.  In principle, for the case of
$B_d\rightarrow \pi^+\pi^-$ the Penguin amplitude has a different
phase than the decay amplitude, and so this decay is not theoretically
pristine.  However, the fact that likely \cite{Witherell}
\[
BR(B_d\rightarrow \pi K)\simeq BR(B_d\rightarrow \pi\pi)
\]
may alleviate this problem, since $B_d\rightarrow \pi K$ is purely
Penguin dominated and the $b\rightarrow s$ Penguin contribution should
be much bigger than the $b\rightarrow d$ Penguin contribution.  So,
effectively, also for $B_d\rightarrow \pi^+\pi^-$ one probably has
only one dominating weak amplitude.  At any rate, as Quinn\cite{Quinn}
discussed, one can in principle disentangle this issue by doing a Dalitz
analysis of the decay $B_d\rightarrow \rho\pi$.  What is more
troubling is that the allowed unitarity triangles permits
the value $\alpha=90^\circ$
(see Fig. 2) and thus the asymmetry in this case is not necessarily
large.  In fact, as Nir\cite{Nir} and others have emphasized, any value for
$\sin~2\alpha$ is consistent with what we presently know about the CKM
matrix.  At a B factory, assuming that $BR(B_d\rightarrow \pi^+\pi^-)
=2\times 10^{-5}$, the relevant error on $\sin~2\alpha$, again with
$100~fb^{-1}$ of data, is about twice as large as that for
$B_d\rightarrow \psi K_S$ ($\delta\sin~2\alpha=0.17$\cite{Abe}).  So
it remains to be seen if one can make a clear measurement of this
angle.

The angle $\gamma$ is even harder to determine.  First of all, to
study it most simply one needs to measure $B_s$ decays and these
will not be accessible at the asymmetric B factories.  However,
prototype decays like $B_s\rightarrow \rho K_S$ do suffer from
``Penguin pollution"\cite{LP} and may not be theoretically pristine
enough.  As Gronau and London\cite{GL} emphasized, it is possible to
extract $\gamma$ by studying $B_d$ decays to non CP self-conjugate
states, like $B_d\rightarrow D^oK^*$.  However, then one is forced
to compare different processes and the prospects for an accurate
determination are not very sanguine\cite{Quinn}.

In my view, measuring a large asymmetry connected with $\sin~2\beta$
in B decays will go a long way towards establishing the validity of
the CKM paradigm.  It is of course true that, accidentally, the rate
difference between $(B_d)_{\rm phys}\rightarrow \psi K_S$ and
$(\bar{B}_d)_{\rm phys}\rightarrow \psi K_S$ could be large in a
non CKM context, but it would be a remarkable coincidence.  This
said, however, as Quinn\cite{Quinn} emphasized here, it is important
to check that the unitarity triangle actually closes.  For instance,
as Branco\cite{Branco} discussed in his talk, it is relatively easy
to change the mixing phase in the $B_d-\bar{B}_d$ mass matrix and
this, effectively, makes the $3\times 3$ CKM matrix non-unitary.
Clearly, much experimental and theoretical work still awaits us before
it can be said that we understand the phase structure in the flavor
sector.  If some discrepancies from the CKM expectations were to be
found in B decays, looking at processes like $B_d\rightarrow K_S K_S$,
which is a pure $b-s$ Penguin decay (and thus should have zero asymmetry)
might be a good diagnostic\cite{Quinn}.

\section{The mysteries}

There were a number of talks in the Conference which addressed important
and still mysterious structural issues in particle physics, like
supersymmetry and supersymmetry breaking and
neutrino masses and mixing.  I have
decided not to attempt to summarize
these talks here because they either were not
quite germane to the main topic of the Conference or they were given too
near to my own concluding talk to sensibly be able to report on them.  I
would like, however, to end by discussing two topics briefly which are
both germane to CP violation and where the issues remain quite open and
challenging.  These concern: baryogenesis at the electroweak scale, which
was addressed here by Shaposhnikov\cite{Shaposhnikov}, Dine\cite{Dine}
and Yanagida\cite{Yanagida}, and the strong CP problem of which
various aspects of it were discussed in the Conference by Vainshtein
\cite{Vainshtein}, Kikuchi\cite{Kikuchi}, Branco\cite{Branco} and
Schierholz\cite{Schierholz}.

There are a number of issues which are widely agreed
upon concerning electroweak baryogenesis and before I try to
outline where the controversies reside, it might be useful if I
summarized these non-controversial points first.  The standard
model possesses 2 of the Sakharov conditions\cite{Sakharov}
necessary for baryogenesis: it violates C and CP and it
violates baryon number, B.  The violation of baryon number in the
standard model is a quantum effect\cite{GTH} and is the result of
a chiral anomaly\cite{Anomaly} in the (B+L)--current.  Since this
current is carried by all the quarks and leptons, there is a selection
rule governing (B+L)--violating processes relating the total change in
this quantity to the number of generations $N_g$ (presumably $N_g=3$):
\[
\Delta(B+L)=2N_g~.
\]

Since $2N_g$ is also the possible amount by which the
topological index of the
electroweak gauge field vacuum configurations
changes by, (B+L)--violating
processes in the standard model involve non-trivial changes in these
configurations.  At zero temperature, these gauge field vacuum
changes are strongly suppressed by a tunneling factor\cite{GTH} and
baryon number violation is vanishingly small
($\Gamma_{\rm B+L~viol.}\sim {\rm exp}~-4\pi/\alpha_W$, with
$\alpha_W=\alpha/\sin^2\theta_W\sim 1/30$).  However in a
temperature bath--like in the early universe--these gauge field
vacuum changes can occur via thermal fluctuations, and the rate
for (B+L)-violation can become important.  This was the crucial
observation of Kuzmin, Rubakov and Shaposhnikov\cite{KRS}.  One
finds, in fact, a rate for (B+L)-violation that is suppressed by a
Boltzmann factor below the temperature of the electroweak phase
transition and is unsuppressed above\cite{Dine}\cite{Shaposhnikov}:
\[
\Gamma_{\rm B+L~viol.}\sim\left\{\begin{array}{ll}
e^{-E_{\rm sph}(T)/T} & T<T_c \\
(\alpha_WT)^4        & T>T_c
\end{array}
    \right.
\]
Here $E_{\rm sph}(T)$ is, roughly, the height of the barrier separating
the two gauge vacua and is related to the electroweak order parameter:
\[
E_{\rm sph}(T) \sim M_W(T)/\alpha_W \sim
\langle\phi(T)\rangle/g_2~.
\]

A comparison of the above rate to the rate of expansion of the
universe, given by the Hubble constant at tempeature
$T~(H\sim T^2/M_{\rm Planck})$, shows that, as a result of
standard model interactions, (B+L)-violating processes are
in equilibrium ($\Gamma_{\rm B+L~viol.}>H$) during a large
temperature interval in the universe:
\[
T^*\sim 10^2 {\rm GeV} \leq T \leq T_{\rm max}\sim 10^{12} {\rm GeV}~.
\]
During this period, any primordial (B+L)-asymmetry in the universe
will be erased\cite{KRS}.  This phenomena has obviously a direct
bearing on the presently observed B-asymmetry in the universe and
suggests two different alternatives:
\begin{enumerate}
\item[i)] The observed B-asymmetry ($\eta_B\sim 4\times 10^{-10}$) is
a result of a primordial (B-L)-asymmetry (or of a primordial
L-asymmetry, as in the model discussed by Yanagida\cite{Yanagida}
at this Conference) which is not affected by standard model
processes.
\item[ii)] There is no primordial (B-L)-asymmetry and since any
primordial (B+L)-asymmetry is erased during the evolution of the
universe, the observed B-asymmetry must be generated at the
electroweak phase transition.
\end{enumerate}

The talks of Dine\cite{Dine} and Shaposhnikov\cite{Shaposhnikov} here
concentrated on this second very intriguing possibility.  To generate
$\eta_B$ at the electroweak phase transition, three conditions need to
hold.  First, the electroweak transition must be first order, so
that the last Sakharov\cite{Sakharov} condition for baryogenesis--that
baryon number violating proceseses must be out of equilibrium--is
satisfied.  However, at the electroweak phase transition one must,
in addition, make sure that this transition is sufficiently {\bf strongly}
to first order, so that the generated asymmetry is not subsequently
erased.  This second condition requires that $\Gamma_{\rm B+L~viol.}
(T^*) < H(T^*)$, which will obtain provided $E_{\rm sph}(T^*)/T^*$ is
sufficiently big.  Numerically\cite{Dine} this
requires
\[
\langle\phi(T^*)\rangle/T^* \geq  1~.
\]
Finally, one must also require that the true-vacuum nucleation
during the phase transition is both sufficiently efficient, and
sufficiently {\bf fermion-antifermion asymmetric} (this is where
CP violation enters into the problem), so that the resulting $\eta_B$
produced is big enough.  This is the most tricky part,
requiring a real calculation of the kinetics of the problem.

It is in these last two points where there is not full agreement
in the literature--a disagreement which was exemplified here by
the two different points of view expressed by Dine\cite{Dine} and
Shaposhnikov\cite{Shaposhnikov}.  Using the finite temperature
effective potential $V_{\rm eff}(\phi,T)$ for the standard model
Higgs sector one can calculate $\langle\phi(T^*)\rangle/T^*$.
As pointed out long ago by Shaposhnikov and collaborators
\cite{BKS}, $\langle\phi(T^*)\rangle/T^*$ will only be large
provided one has a light Higgs boson in the theory.  Conversely,
using the bound on the Higgs mass provided by LEP
($M_H > 62.5$ GeV\cite{LEP8}), from a calculation of $V_{\rm eff}(\phi,T)$
one can obtain an upper bound for $\langle\phi(T^*)\rangle/T^*$.
Recent calculations of $V_{\rm eff}(\phi,T)$\cite{DHLL} for the case
of one Higgs doublet using the LEP bound
give $\langle\phi^*(T)\rangle/T^* \leq 0.5$.
This led Dine\cite{Dine} to conclude that one cannot generate the
baryon asymmetry at the electroweak scale in this simplest model of
symmetry breakdown.

Shaposhnikov\cite{Shaposhnikov} disagreed
with this conclusion
for two reasons.  First, he did not trust the calculation of
$V_{eff}(\phi,T)$ because of infrared problems.  Second, he believes
that for this dynamical problem it does not suffice to compute
$\langle\phi(T^*)\rangle/T^*$ via an effective potential, but one
must really compute directly $\Gamma_{\rm B+L~viol.}(T^*)$.  I
believe the first objection is not that germane, at least for the
region of Higgs masses in question.  However, the second point
may well be relevant. In this case it may be premature to imagine
that there is a conflict between electroweak baryogenesis and a one Higgs
doublet model of symmetry breaking.

The second area of disagreement between Dine and Shaposhnikov is more
technical and is connected with the actual mechanism by which a
fermion number asymmetry is generated at the electroweak phase
transition.  Roughly speaking, as a bubble of true-vacuum nucleates
during the phase transition, the bubble wall because of CP violating
interactions will have different transmission coefficients for
fermions and antifermions.  The net baryon asymmetry $\eta_B$ is
proportional to the difference in transmission rates for baryons
and antibaryons at the bubble wall.  Dine\cite{Dine} argued that
this rate difference in the CKM model--as in the vacuum--is
irrelevantly small (of $O(10^{-22})$) because it is totally GIM
suppressed.  One finds:

\begin{eqnarray*}
\eta_B & \sim & \left[(m_t^2-m_c^2)(m_t^2-m_u^2)(m_c^2-m_u^2) \right.\\
& \times & (m_b^2-m_s^2)
(m_b^2-m_d^2)(m_s^2-m_d^2) \\
 & \times & \left. s_{12} s_{23}s_{13} \sin\delta \right]\frac{1}{M^{12}_W}~.
\end{eqnarray*}

Shaposhnikov\cite{Shaposhnikov}, on the other hand, argues that
in a thermal bath the GIM factor is largely erased, so that in
the standard model one can in fact obtain a sizeable asymmetry.
Although I cannot really judge whether the calculation of
$\eta_B$ presented by Shaposhnikov (based on the work of Farrar
and Shaposhnikov \cite{Farrar}) is reliable, it seems likely to
me that thermal effects can help ameliorate the GIM suppression
one naively calculates.

My own conclusion regarding electroweak baryogenesis is that it
is unlikely that it
actually proceeds in the simplest model of symmetry
breaking.  With one Higgs doublet, the electroweak phase transition
is probably not sufficiently strongly first order to prevent
sizable erasure of the produced asymmetry. Furthermore, this asymmetry is
already probably too small due to having too simple a source for
CP violation.  However, in extended Higgs models, both of these
difficulties are ameliorated.  Hence, if the baryon asymmetry in
the universe was produced at the electroweak scale, one should
expect to have other sources of CP violation, besides the CKM phase
be relevant at low
energies.  Thus electroweak baryogenesis and an edm in the
$10^{-26}$ ecm range may well be closely connected!

Let me end by making a few remarks on the strong CP problem.  The
presence of the term
\[
{\cal{L}}_{\rm strong~CP}= \bar{\theta}
\frac{g_3^2}{32\pi^2} F_a^{\mu\nu} \tilde{F}_{a\mu\nu}~,
\]
with $\bar{\theta}$ being the sum of the QCD vacuum angle $\theta$ and
the contribution from the quark mass matrix
\[
\bar{\theta} = \theta + \arg \det M~,
\]
follows from a careful examination of the correct vacuum
structure of QCD\cite{CDG}.  Although one can try to avoid the
appearance of the strong CP interaction by modifying the $|\theta\!>$-
vacuum structure of QCD, as it has been suggested recently by
Samuel\cite{Samuel}, one then gets back into trouble with the old U(1)
problem that $m_\eta \not= \sqrt{3}~m_\pi$!  This is basically the
criticism presented here by Kikuchi\cite{Kikuchi} to the ``solution"
to the strong CP problem of Samuel.  The renormalization properties of
$\bar{\theta}$ discussed in the talk by Vainshtein\cite{Vainshtein} made
it apparent that in all aspects this parameter acts precisely as would
the coefficient of any dimension 4 operator.  The concomitant GIM
factors which entered in Vainshtein's discussion are also totally
natural, since as any quark becomes massless it is necessary that
$\bar\theta\rightarrow 0$\cite{JR}.

Unfortunately, these comments do not bring one closer to a solution
to the strong CP problem.  The present bounds on the neutron edm
require that $\bar\theta < 10^{-9}-10^{-10}$\cite{Baluni} and there
is no immediate explanation why this effective angle should be so
small.  Of course, I am still prejudiced towards the solution to
this problem which I proposed with Helen Quinn some time ago
\cite{PQ}!  However, one has no proof as yet for the presence of
a $U(1)_{PQ}$ global symmetry since no axions (visible or invisible)
have been found.  In my view, the PQ solution is preferable to
solutions of the strong CP problem involving soft CP breaking, since
models of this type run into a variety of difficulties unless one is
very careful.  This was nicely illustrated
at this Conference by the example discussed
by Branco\cite{Branco}.

The difficulties which soft CP breaking models
encounter are different depending
on whether the breaking of CP occurs at low scales (i.e. near the
electroweak scale) or at high scales (i.e. near a GUT scale).  In the
first case, as in the model Branco discussed, one has
problems with the enormous energy density that resides in the walls
separating domains in the universe, which vastly exceeds
the universe's closure density\cite{KOZ}.  If CP is broken
spontaneously at high scales, one can avoid this domain wall problem
through inflation.  However, then it is difficult to transmit the CP
violating phase generated at the high scale to the low energy sector
\cite{Blois}.  The most successful models of this type have been
devised by Nelson and by Barr\cite{BN} and, in general, tend to be
of a superweak variety.  Therefore, if these ideas are correct one
would expect quite different predictions for CP violation in the
B system!

These brief comments on electroweak baryogenesis and on the strong CP
problem emphasize an important lesson, which is worthwhile keeping
in mind.  Namely, that even though the elucidation of some of the
deep mysteries may perhaps escape us theoretically at the moment, new
experimental data can have a profound impact on our understanding.
Indeed, new data may expose our present beliefs only as theoetical
prejudices.  Let us hope that future experiments probing for CP
violation may have this salutary effect!

\section{Acknowledgements.}

I am grateful to H. Sugawara and K. Hagiwara for their hospitality
at KEK during this very stimulating Conference. I am also indebted
to the KEK Laboratory and the joint NSF Japan-USA Collaborative program
for travel support.  This work was supported in part by the
Department of Energy under Grant No. FG03-91ER40662.

\end{document}